# Data Analytics on Online Labor Markets: Opportunities and Challenges

Michael Feldman, Frida Juldaschewa and Abraham Bernstein

The data-driven economy has led to a significant shortage of data scientists. To address this shortage, this study explores the prospects of outsourcing data analysis tasks to freelancers available on online labor markets (OLMs) by identifying the essential factors for this endeavor. Specifically, we explore the skills required from freelancers, collect information about the skills present on major OLMs, and identify the main hurdles for out-/crowd-sourcing data analysis. Adopting a sequential mixed-method approach, we interviewed 20 data scientists and subsequently surveyed 80 respondents from OLMs. Besides confirming the need for expected skills such as technical/mathematical capabilities, it also identifies less known ones such as domain understanding, an eye for aesthetic data visualization, good communication skills, and a natural understanding of the possibilities/limitations of data analysis in general. Finally, it elucidates obstacles for crowdsourcing like the communication overhead, knowledge gaps, quality assurance, and data confidentiality, which need to be mitigated.

## Introduction

In the past years it has become evident, that there is a continuously growing demand for data scientists and for people who are able to systematically interpret data. Since the availability of data is growing faster than the availability of experts with the relevant skillset to interpret it, finding competent experts for data analysis tasks is becoming increasingly challenging due to a variety of required skills. Contrary to the past, when benefiting from data products was a prerogative of big companies, having server farms and in-house teams of support technicians, currently, the development of cloud computing allows for on-demand usage of computational resources at reasonable costs. Therefore, even small and medium sized enterprises (SMEs) have started collecting data about their customers, business transactions, and other records related to their business. However, analysis of the collected data is hampered by the growing shortage of data experts capable of analyzing the data and producing comprehensive insights that are intelligible to a wide audience of population and decision makers (Davenport and Patil 2012).

Often, lacking internally available talent, companies are compelled to seek for external solutions that will allow to make sense of their data. Recognizing the need for on-demand data analysis, multiple software companies, such as Microsoft and Google, started developing and offering cloud-based data analytics products that allow running various machine learning (ML) algorithms on a big scale in the cloud. Some of these products claim to reduce the complexity threshold of data analysis by allowing to plug the data to their services and subsequently to run data analysis as a black box process. However, this approach leads to questionable results given the limited control over the data analysis process and reduced flexibility to preprocess the data and to tailor the models for specific needs. In addition, statistical expert systems (Serban et al. 2013) have been proposed to address the matter. Many of these systems require data analysts in order to be employed correctly. Therefore, even though data analysis has been made much more accessible by this variety of tools, there is an apparent need for skilled experts to conduct and orchestrate the process of data analysis.

The lack of experts available in geographical proximity can be resolved by online labor markets (OLM), that overcome multiple drawbacks and allow to hire experts in a flexible manner. The potential of such platforms has been recognized by the industry and the earnings of such platforms experienced sharp increase during past years, attracting growing numbers of freelancers and employers. The recruitment process is crucial to the success of OLM and potentially, it could be supported by the substantial research body on the personnel selection theory that is well known in organization psychology. However, the nature of recruitment in OLM is different from traditional candidate selection due to its remote character and high level of uncertainty with regard to workers' skills and abilities. Unfortunately, research on the necessary skills for potentially crowdsourced data analysis still remains in a blind spot and has not been sufficiently addressed yet. The IS community has investigated the interdependencies between tasks and different psychological and environmental aspects of candidate individuals. However, in the OLM context, so far, no theoretical frameworks have been proposed to describe employer-employee interactions in the candidate selection process. In our study we rely on the Person-Job (P-J) fit framework (Edwards 1991) to understand how requirements for data experts match the existing talent on freelance platforms. P-J outlines the importance of the needs-supplies conceptualization as an important factor for good fit. For the highly important field of OLM for data analysis such a conceptualization is missing. Therefore, we carry out a study to both analyze the requirements and explore the skills available online. We argue for the importance of this phenomenon and hope that our work will promote the discussion on the major constructs composing the candidate selection process on OLM. Consequently, *our work explores the explicit factors in online employee selection by adopting the needs-supplies dimension of the Person-Job theory.*

Various data analysis tasks require diverse knowledge and skills. While some tasks are fairly straightforward and effortless, others require proficiency in multiple disciplines and hands-on experience. To illustrate, descriptive statistics require a somewhat shallow statistical background while other methods, such as neural networks or support vector machine learning frequently require in-depth understanding. Therefore, our first research question is: *RQ1 What are the skills required in data analysis?* Identifying the skills required for a certain task is the first step in finding suitable workers to perform it. For the successful assignment of freelancers to tasks we also need to understand the capabilities of the workers. Hence, gaining an insight about the skills of freelancers will allow to design tasks while taking into account the constraints imposed by the existing talent pool in the online labor market. Therefore, the second research question is: *RQ2 What are the relevant skills and characteristics that freelancers in OLM possess, and do they match the required skills for data analysis (identified for RQ1)?* The last research question deals with various obstacles to the outsourcing of data analysis to freelancers and gain insights on the potential ways to resolve these problems. Hence, our third research question is: *RQ3 What are the obstacles to outsourcing data analysis to OLM?*

To answer these questions, we conduct an exploratory study including 20 interviews with data scientists, followed by a survey with 80 freelancers to learn about the talent available on major freelance platforms. Our contribution is twofold: 1) we fill the research gap concerning the endeavor of outsourcing data analysis to OLM by means of systematic research on needs vs. supplies of explicit skills and 2) provide the necessary basis for future theorization on employee selection in an online setting.

The remainder of this paper is structured as follows: In the next section we explore the related work in the domains of crowdsourcing as well as online labor markets and give an overview of the theoretical work related to our study. Then we describe our research approach and outline the qualitative and quantitative results of our study. Next, we discuss the results of our study and present limitations and future research. Finally, we conclude with a summary of the paper.

## Related Work

### The Crowdsourcing Phenomenon and Analytics

Crowdsourcing has gained increased relevance and acceptance as an approach for outsourcing activities to an online community. In recent years, the business community embraced the approach of outsourcing some activities to a crowd by means of evolved and specialized web-based platforms. Jobs are mostly partitioned into groups of simplified sub-tasks and distributed to crowd workers in an open call manner (Howe 2008; Alam and Campbell 2014). Even though crowdsourcing has a long history (just consider the Longitude prize or the Oxford English Dictionary), its current popularity is largely accounted to Amazon's establishment of the first crowdsourcing Internet marketplace called Mechanical Turk (MTurk).[1] This platform provides a wealth of paid micro-tasks that require minimal time and cognitive effort, but aggregates results in major accomplishments (Kamar et al. 2012). As for today, MTurk has preserved its leading role as the most popular platform for micro-tasks. However, the phenomenon of online work has expanded to include platforms supporting laborious expert work in various domains. Platforms such as Upwork and Freelancer,[2] offer the service of matchmaking between employers and freelancers based on reported expertise. Other platforms such as TopCoder, 99designs, or Kaggle[3] offer contest-based participation, while yet other platforms like InnoCentive or Idea Bounty[4] are crowdsourcing ideas for solving challenging problems. Encouraged by the potential of human computation, recently, attempts have been made to extend human computation tasks beyond relatively simple and non-demanding ones (e.g. Haas et al. 2015). As a consequence, the crowdsourcing domain is faced with an emerging need for concepts and paradigms for the assignment of complex tasks that require a wide spectrum of human abilities and talents to suitable crowd workers. Finding appropriate crowd workers, is non-trivial due to the motivational, cognitive, and error diversity of humans (Bernstein et al. 2012). Moreover, the remote and unstable character of most crowd markets limits the ability to track and profile workers and gives rise to an even greater challenge of establishing trustworthy and robust recruiting policies (Ipeirotis 2010).

Evidently, platforms such as MTurk are primarily designed for micro-tasks and do not support complex and ill-defined tasks that require well-established coordination and trust relationships between the requesters and crowd workers. The rise of freelancer platforms aims to fill this gap by supporting both, employers and crowd workers/freelancers, with a user interface and workflow that enables to conduct complex projects in a remote manner. In contrast to the micro-task labor market that has reached saturation and even some decline (see Ipeirotis' analysis on MTurk[5]), the online freelancer market has grown substantially during the last years. It is now bringing together millions of freelancers and employers (Agrawal et al. 2013). This shift exemplifies the transition of online labor markets from simple, short-term, and low effort jobs, as they were originally common in online labor markets, to complex and long-term tasks that are typical to traditional work settings. A growing number of white collar workers are switching to online labor markets due to the advantages online work can offer. Students, homemakers, retired experts, and single parents are frequently found on OLM and the prevailing professions are graphic design, copywriting, data entry, and programming (Mill 2011).

---

[1] www.mturk.com/mturk/welcome
[2] www.upwork.com, www.freelancer.com
[3] www.topcoder.com, www.99designs.com, www.kaggle.com
[4] www.innocentive.com, www.ideabounty.com

Similarly to the general shortage of data scientists, i.e. experts able to provide comprehensive data-driven solutions, (Davenport and Patil 2012), it is fairly uncommon to find data analysis experts on a crowdsourcing platform. It is, however, not unusual to find workers possessing some partial knowledge and/or willing to learn a new topic. A survey of 153 data scientists, conducted by Crowdflower (2015), has revealed that data scientists mostly refer to themselves as researchers (54%), computer scientists (52%), BI analysts (36%), and mathematicians (19%). Additionally, most of the respondents mentioned they are working with Excel (56%), R (43%), and Tableau (26%). The majority of the respondents consider data cleaning and organizing as the most time consuming task (67%), while 53% say that collecting data sets is the most laborious. This matches with Kurgan and Musilek (2006) who surveyed multiple papers evaluating the relative effort in different data analysis activities and concluded that data preparation is by far the most time consuming activity with estimates ranging between 45 and 60%. The most demanded skills are programming and statistics and proficiency in Python and R are by far the most prominent. According to Harris et al. (2013), a data scientist has to be capable of 1) designing statistical models, 2) creating machine learning and text mining algorithms, 3) cleaning and converting raw data, 4) carrying out quality assurance testing to ensure the quality of models, and 5) communicating the results through clear data visualization. Other supplementary skills such as communication, collaboration, and creativity are also mentioned as key success factors. Chatfield et al. (2014) have analyzed a body of literature from six major academic databases and derived a set of six data science attributes: 1) Entrepreneurship and business domain knowledge, 2) Computer scientist, 3) Effective Communication skills, 4) Create valuable and actionable insights, 5) Inquisitive and curious, and 6) Statistics and modeling.

The aforementioned studies provide a high level understanding of who online freelancers are and what data analysis tasks that can be outsourced. They, however, do not shed light on what concrete, well-specified skills are required and whether they are available on OLM.

## *Theoretical Background*

In this section we first discuss an overview of existing research on hiring policies common in online markets and then review theoretical frameworks relevant to the online matchmaking of freelancers with employers. We then support our choice to adopt the Person-Job fit theory as a theoretical lens for our study.

Evaluating skills of job candidates is one of the major challenges in both online and traditional labor markets. Even though some platforms allow job candidates to perform various online tests to assess their competence in a variety of topics, cheating and tests' leakage hamper reliable evaluation. Moreover, technological advancement requires tests to be frequently updated and reliably evaluated, promising that the performance on these tests adequately reflects a candidate's skills (Christoforaki and Ipeirotis 2015). One major difference of online compared to traditional labor markets is the highly heterogeneous workforce composed of a crowd with different skills spanning across a variety of different professions. While the candidates in traditional markets share some similarities such as common cultural and geographical background, candidates in online labor markets come from all around the world and exhibit high variance in qualities and skills. Kokkodis and Ipeirotis (2015) assume these skills to be latent, however, possible to be measured through the worker's available characteristics on a platform such as employee ratings, accomplished projects, hours worked, and wages. Utilizing these characteristics, they present a number of models that estimate the workers' latent skills and their evolution over time. Feldman

---

and Bernstein (2014) propose that cognitive abilities of freelancers are another latent factor that, to a large extent, predefines the performance on various crowd-tasks. They examine the performance on various crowd-tasks with different setups to predict task performance where cognitive abilities, performance on previous crowd-tasks, or both are partially known, and show that cognition-based task assignment leads to an improvement in task performance prediction. Suzuki et al. (2016) propose to support the skill development of workers by introducing a concept of micro-internships. According to the proposed solution, micro-internships allow workers to learn, improve, and develop new skills. At the same time employers can evaluate the skills of a candidate. Verroios et al. (2015) are grouping employers on Odesk (today Upwork) with respect to their hiring criteria and learn the hiring preferences for each cluster. Results show that while some groups of clients are positively biased to freelancers that are new to a marketplace, others ignore their reputation and focus primarily on a person's job fit. In this context, Pallais (2013) uses a field experiment in Odesk to show that awarding new freelancers with a first job benefits the market with information about their abilities and increases freelancers' average earnings.

The theoretical foundations of our exploratory study can be seen in the Person-Job fit element of the well-studied Person-Environment fit theory (P-E). The concept of P-E fit originates from the interactionist theory of behavior (Chatman 1989). This theory proposes that neither personal nor the environment characteristics on their own can fully explain the compatibility between people and organizations, but rather the interaction between these two yield a better understanding of the fit between the person and the environment. P-E has three common conceptualizations: 1) *Supplementary vs. complementary fit*: while supplementary fit occurs when persons possess characteristics that are similar to the other individuals in the environment, complementary fit arises when the traits and skills possessed by the individual complete the missing characteristics. 2) Further, the complementary fit can be distinguished between *needs-supplies vs. demands-abilities fit.* Needs-supplies fit occurs when an organization satisfies an individual's needs, desires, or preferences. On the other hand, the demands-abilities perspective suggests that fit occurs when an individual has the abilities required to meet organizational demands. 3) The last conceptualization is *objective vs. subjective fit*. Subjective (or perceived) fit is conceptualized as the assessment whether an individual fits well in the environment, while objective (or actual) fit is the congruence between the person and the environment that is externally measured with indirect measures such as outside evaluators (Kristof 1996; Sekiguchi and Huber 2011).

Person-Job (P-J) fit, one of the main constituents of Person-Environment fit, is in the core of employee selection studies and is primarily concerned with finding suitable candidates by analyzing the demands of the job and assessing candidates' skills and abilities. Good P-J fit leads to high job proficiency and, therefore, work is likely to be accomplished quickly and with higher quality than in circumstances when the job proficiency is low (Werbel and Johnson 2001). Originating with fairly simple interpretation sparked by Taylor's monograph on the principles of scientific management (1911), the process has gained sophistication with the formulation of statistically reliable processes that can be used to determine P-J fit (Werbel & Gilliland 1999; Brkich et al. 2002). The main conceptualizations relevant to P-J fit are *needs-supplies* and *demands-abilities,* and, as discussed earlier, both are extended conceptualization of *complementary fit*. The needs-supplies conceptualization incorporates the desires of individuals, and the characteristics and attributes of the job that may satisfy those desires. The desires can be seen as a mixture of goals, psychological needs, interests, and values attributed to the person, whereas the job supplies are described as general characteristics of pay, occupation, and job attributes. The demands-abilities perspective consists, on the one hand, of the job demands that are required in order to carry out the tasks of the job and, on the other hand, of the abilities that the individual has and can use to meet the job requirements. Job demands typically consist of the knowledge, skills, and abilities required to perform at an acceptable level in the job. Abilities include education, experience, knowledge, and skills. In employee selection practices, strategies to assess P-J fit include resumes, tests, interviews, reference checks, and a variety of other selection tools (Sekiguchi 2004). There is considerable evidence that a high level of P-J fit has a number of positive outcomes. The review of the P-J fit literature by Edwards (1991) identified job satisfaction, low job stress, motivation, performance, attendance, and retention as outcomes that are positively affected by P-J fit. When P-J fit is assessed as the overlap between an individual's desires and benefits received from performing a job, it leads to improved job satisfaction, commitment, and reduced intentions to quit. *The goal of our study is to identify the important elements necessary for the needs-supplies dimension of the Person-Job theory in OLM for data analysis.*

# Research Approach

Our study adopts a sequential mixed-method approach, harnessing both the power of qualitative as well as of quantitative research (Tashakkori and Teddlie 2009). The qualitative study comprises an interpretive case study, in which 20 data analysis experts are interviewed (Walsham 2006). The quantitative study consists of a web survey (Dillman 2011) following a descriptive and cross-sectional research design as it has been outlined by Pinsonneault and Kraemer (1993). The results of the first research phase inform the design of the second research phase. Using such a multiple-method approach leads to higher robustness of results due to triangulation – leveraging the usage of multiple methods, data sources, or theories to facilitate deeper understanding of the phenomenon (Denzin 1973). Our approach allows on the one hand for data triangulation, the use of a variety of sources in a study, and on the other hand for methodological triangulation, the use of multiple methods to study a research problem. As qualitative research is especially appropriate for studying complex phenomena, we applied this method first to explore the domain of data scientists and to gain in-depth descriptions and understanding of their environment. After having gained a thorough understanding, quantitative research is well suited to apply the learned content onto a broader population and obtain quantitative data to generalize research findings (Johnson and Onwuegbuzie 2004). Overall, our process consists of four steps: first, qualitative data is collected through interviews with data scientists and data analysts. Second, the qualitative data is used to identify necessary skills and other factors for outsourcing data analysis tasks to freelancers. Third, an online survey is designed based on the previously identified skills and distributed on various freelancing platforms. Lastly, the generated quantitative data is analyzed to compare the skills available on online labor markets with the desired skills.

## Qualitative Research

We approached potential participants for interviews through personal contacts, professional online business networks (e.g. LinkedIn, XING, Data Science Central),[6] and professional meetups[7] in our city.[8] By exploring various sources, we intended to compose a sample of individuals with diverse backgrounds, spanning different industries and positions. The main prerequisite for participation in our study was the profession criterion; specifically, we aimed at individuals who either hold positions of data scientists or data analysts, or are primarily occupied with data analysis in their daily work. Whereas job titles and job requirements vary throughout organizations, the fact that "a data scientist represents an evolution from the business or data analyst role" (Zikopoulos et al. 2012) suggests that they have a common foundation and work with data to answer business questions (Kandel et al. 2012). Therefore, all selected participants are experts in data analysis, and thus able to provide valuable insights about the domain of data analysis. In total, 20 semi-structured interviews, lasting between 30 and 60 minutes, were conducted during a seven-week period in December 2015 and January 2016, with the researcher taking the role of an outside observer (Walsham 1995). Based on the participant's preference, 14 interviews were conducted in German and 6 in English. The interviewer was fully proficient in both languages.[9] This follows Myers and Newman's (2007) suggestion to create a friendly environment, noting the importance of interviewees being able to use their own language, which increases the likelihood of disclosure. Due to geographical and time limitations of some participants, two interviews were conducted via Skype, while the others were performed face-to-face in locations preferred by the participants. All interviews were recorded and then transcribed. Also, notes were taken during the interviews to capture complementary non-verbal insights.

Overall, we closely followed the principles proposed by Klein and Myers (1999) as well as those of Myers and Newman (2007). To ensure the coverage of all important questions, we conducted semi-structured interviews using a question script, which allowed for free development of the dialogue and assured a similar structure between all interviews. The interviewer first introduced herself, explained the purpose of the research, and guaranteed confidentiality to the interviewees (Walsham 2006). Then, to begin with the questions, interviewees were asked about their occupational background, their experience with data analysis, typical tasks in analysis projects they often perform, and the skills and tools required for these tasks. Progressing to the main topic of the interviews, participants were asked what tasks they would outsource to freelancers and what kind of skills as well as knowledge freelancers would need to perform these tasks. Finally, we discussed difficulties and few other general conditions in the endeavor of outsourcing data analysis to freelancers on OLM. All questions were open-ended, i.e. designed in a way that participants could specify their own answers without influence of the interviewer, and thus minimized the possibility of biases occurring in the data collection process (Fontana and Frey 2000). Closing the conversation, the researcher offered participants to receive the study results upon completion. After all interview records were transcribed, the available data was iteratively analyzed using coding technique (Miles and Huberman 1984). Following this technique, we first applied open coding where the entire data was explored and broken apart to create codes. Subsequently, applying axial coding, we identified possible connections between codes and concepts (Corbin and Strauss 2008). This iterative process implied repeated examination of the interview data which gradually led to the elaboration of generalizations, i.e. factors and skills necessary for outsourcing data analysis tasks to online labor markets. Eventually, as no new insights were developed after 15 interviews, we concluded that data saturation has been reached and stopped interviewing after 20 interviews (Guest et al. 2006).

---

[6] www.linkedin.com, www.xing.com, www.datasciencecentral.com

[7] www.meetup.com

[8] Name of city hidden for double-blind review.

[9] Quotations taken from the German interviews and used in the Results Chapter were faithfully translated into English language.

### Quantitative Research

After analyzing all interview transcripts, we developed an online questionnaire based on the results of the conducted interviews. To ensure traceability of results throughout the entire research project, and thus also support credibility, consistent term descriptions were used during all the research stages (Cronholm and Hjalmarsson 2011). Following a descriptive and cross-sectional research design (Pinsonneault and Kraemer 1993), the purpose of the questionnaire was to study the distribution of skills, expertise, and knowledge in the population of the most prominent freelance platforms. The cross-sectional design we adopted, implies that data was collected once and, thus, represents the population at that one point in time. We selected freelance platforms based on their size and on the availability of freelancers with data science or data analysis experience. After a thorough analysis of currently available freelance platforms, we chose Upwork and Freelancer as they constitute the biggest online workforce to date in general (~70%) and a large pool of freelancers specializing in different kinds of data analysis. For each platform we received 40 reliable survey submissions that have passed the quality assurance checks, making it a total of 80 (valid) participants, each of which was rewarded with 5 US dollars.

Throughout our survey we followed guidelines provided by Dillman (2011) and Fowler (2013). After the questionnaire was fully designed, it was iteratively pilot-tested with seven respondents. Short discussions with each person led to improvements and helped to refine the final version of the web survey. The final survey[10] consisted of 29 questions, spanning a mixture of primarily Likert-scale style questions as well as few open-ended, multiple-choice, and single-choice questions. The Likert-scale style questions were designated to capture the freelancers' skills, knowledge, and expertise with various tools, programming languages, and statistical methods. These questions were grouped into several matrices, had five-point response scales, and were all constructed in a similar manner (Dawes 2008). Other questions aimed at learning about the respondents' demographics, educational and occupational background, and experience with data analysis. In order to compare the opinions of interviewees and freelancers, we asked them which tasks of a data analysis project they could or could not imagine being outsourced to freelancers on OLM, and what difficulties they foresee in this undertaking. Since surveys, particularly online surveys, are subject to careless or inattentive responses (Meade and Craig 2012), we integrated several quality assurance questions to assure reliability of the collected data (Kittur et al. 2008; Gadiraju et al. 2015). As a result, 10 out of 90 submissions were excluded from the analysis. According to the findings of de Winter and Dodou (2010), the remaining 80 submissions were analyzed by performing one-sample two-tailed t-tests (which was found to perform comparably to the Mann-Whitney-Wilcoxon for the

---

[10] URL suppressed to ensure double-blind review.

5 point Likert scale) in order to identify statistically significant skills and expertise in tools and statistical methods. Moreover, for further analyses, we conducted descriptive statistics analysis and calculated Spearman rank sum correlations.

## Results

In this section we first present the qualitative results of the interpretive case study (i.e. the interviews), and then outline the quantitative results of the web survey study.

### *Analysis of Qualitative Data*

Our 20 interviewees comprised a diverse group of individuals that hold various positions and deal with data science or data analysis. Their experience varied between 4 and 22 years, with a median of 6.5 years. By having such a diverse range of professionals, we hoped to avoid elite bias, misrepresentation, and non-generalizable responses (Miles et al. 2013). The interviewees represent various industries, including insurance (4), financial service (4), digital analytics (5), analytics consulting (3), telecommunications provider (1), retail (2), and Internet broadcasting (1). The interviewees are all experts in data analysis and 95% of them stated that they have acquired their knowledge during university studies where 6 studied Economics, 3 studied Statistics, 3 studied Physics, and the rest studied other areas comprising mathematics and computer science classes. The majority, 60%, holds a master's degree, 25% hold a doctorate degree, and 15% have a bachelor's degree. Seventy percent state to have continued to learn on the job while 35% additionally have made use of online courses, books, or individual programming tasks to improve their skills.

**Skills and Knowledge Required from Data Scientists (RQ1)**

Although the main focus of the interviews was to extract the necessary skillset for freelancers, in order to identify communalities, we first asked interviewees to name the skills they need in their jobs. All interviewees declare technical skills to be absolutely necessary. These include **programming skills** (mentioned by 75% of the interviewees), **database skills** (55%), **machine learning skills** (30%), and general IT affinity like understanding how servers, software, and apps work (30%). Also **statistical and mathematical skills** are mentioned by almost every interviewee (90%), which includes the ability to conduct statistical analyses with various methods/tools and to have a general flair for numbers. P5[11] explained the necessity for statistical and mathematical skills as follows: "*Normally when we talk to stakeholders [...], they don't really agree or [...] understand why you need a big mathematics skill set, but I would argue [...] you still need a good mathematical background to make your conscious decision of the techniques you're using.*" Furthermore, **domain knowledge** (55%) in the field where the data analysis is conducted is considered to be of importance. It is necessary to understand the context and to know the company's business goals, as this influences the direction of an analysis project. Interdisciplinarity is mentioned to be important in this context, as data scientists often have to tackle problems from diverse areas of a company. P14 stated: "*You need to have a pretty broad understanding, not always super deep, [...] to understand what you are actually analyzing.*" Another very important skill is **communication** (55%), i.e. the ability to communicate with different groups of interest and present results in a clear and simple way. If communication fails, the effort to analyze data can be in vain, as P3 stated: "*It doesn't help to just sit at the computer, you can be the best programmer, but at the end you need to be able to communicate the results clearly, and in the beginning you need to understand, what is important for the other person.*" Hand in hand with communication skills goes the flair for **consulting and mentoring** (25%), including working well with customers and keeping an open mind towards their needs. Having a **data-oriented mindset** (45%), i.e. to understand data and its structure, to be sensitive to the alternatives and limitations of data analysis, is another aspect interviewees mentioned several times. This skill of data understanding intersects with the understanding that data scientists need **experience** (35%) in their job in order to be successful (e.g. to know what is the best method to use in each case and how to interact with the data at hand). Moreover, data scientists need to have **logical thinking and reasoning** (45%), such that they are able to structure problems and to break them down, abstract, and operationalize solution steps. As stated, for instance, by P19: "*You need a structured approach, because if you're not a tiny bit structured, you start losing yourself in the data. So you should really always keep in mind, where do we want to go or which variable or function we want to optimize. And then anything we do, we target towards this goal. Otherwise we end up basically analyzing data which is not relevant for us.*" Oftentimes, data scientists are confronted with new problems in a new field, or have to apply new methods, tools, or software. Thus, they need to be **willing to learn** (20%) and to keep a **curious attitude** (20%). Curiosity is not only mentioned in connection with learning new things, but also as an aspect of being curious about what can be found in the available data. As stated by P16: "*This analytical curiosity, that you can simply recognize certain structures in the data itself, or also exercise patience to play around a little bit and to look in what direction the whole issue will develop.*" Also, data scientists and analysts need to be patient and enduring, as data analysis projects often require exhausting examination of data. P4 says: "*Sometimes you spend hours on some tools or some data and you don't find anything. So it's very hard sometimes to keep you motivated.*" Thinking about the problem deeply and trying to get to the bottom of it is therefore an important trait of data analysts. This includes paying attention to details and having an intuitive and inventive mind. This was mentioned by several interviewees (25%), e.g. by P7: "*You need to learn to pay attention to the last, to the smallest detail. You will, and that's guaranteed, stumble over those at the end.*" At the same time, they need to be skilled in **project management** (15%) as data analysis projects require adhering to constrained timelines. P10 highlighted: "*Oftentimes you work under time pressure. [...] You oftentimes also don't get the data that you need at the right time. So sometimes you have to prepare scripts blindly, then you receive the data and apply the scripts onto the raw data.*" Several interviewees (30%) note the fact that a data scientist cannot have all skills. Data science or analysis projects require **teamwork** and thus each analyst has a specific role with a corresponding skill set: "*Usually you just pick one or two max, that you try to perfect in a sense*" (P4). This skill set can be composed of the previously discussed skills. As to the software, a great majority of the interviewees work with R (85%) and Python (75%) in their jobs. In general, the opinion prevails that one programming language is required but that it is irrelevant which one. Rather, it is important to "*having acquired all the programming logic in one or another language*" (P2). Also SQL (55%) and Excel (45%) are commonly used among the interviewees. Other tools mentioned several times include Java, JavaScript, Tableau, Hadoop-related technologies, ETL tools, and Oracle. Regarding statistical methods, P11 said: "*It's a full zoo of methods, over which you need to have a little overview.*" Mainly, interviewees utilize both descriptive statistics and data mining methods such as regressions, clustering, classifications, predictions and a number of machine learning algorithms. P16 looked at this topic with sorrow: "*90% of our jobs are descriptive. And all the cool stuff, that is really fun, is unfortunately done way too rarely.*"

---

[11] For the purpose of reporting we numbered the interviewees from P1 to P20.

**Skills and Knowledge Required from Freelancers (RQ1 continued)**

Since we are interested in the entire skill set that should be present on a freelancing platform, we asked interviewees for the desired skills of freelancers for different tasks, and combined all answers to receive a full overview of necessary skills. **Statistical and mathematical skills** as well as **database skills** were mentioned most necessary by the interviewees (80% each). Next, **programming skills** (65%) are mentioned to be important. Specifically, almost every interviewee highlighted freelancers should know R (90%), followed by Python (65%). In general, similarly to the responses about their own skills, they regarded one programming language as necessary for freelancers without preferring one over the other (40%): "*I would say that it doesn't matter which programming language, because someone who knows one programming language, learns a new one very quickly" (P13).* Furthermore, freelancers should be familiar with SQL (45%), Excel (40%), ETL tools such as Talend (25%), and several other tools, data formats, and operating systems (mentioned by up to 20%). Also **domain knowledge** is mentioned by 65%. P20 explained this necessity as follows: "*Data analysis per se doesn't exist. It's always data analysis in a context: logistics, medicine etc. So the know-how in this context, in this domain, is absolutely necessary [...]. You need to explain first what the data means [...]. If people don't understand it, they cannot identify the data quality problems etc.*". **Data understanding**, the ability to understand data and its structure and how to work with it, was also identified as an important skill (50%). Forty-five percent mentioned **communication skills** that encompass being able to talk to different groups of interest, being an "*interface person*" as called by P7, and communicate results in a clear and straightforward way. P3 stresses the importance of communication: "*I think when it's external, it's even more important that there* is good communication, because the person doesn't know the company well and the company doesn't know the person well. That's why you have to pay attention to communication all the more." Furthermore important are **visualization skills** (40%) (i.e. the ability to visualize data in a meaningful way), having an eye for simple and clear design, and attention to details, e.g., making sure *"that you don't make it red and green, but maybe orange and blue, because people have a red-green deficiency"* (P14). Having **experience** (30%) is an advantage, as it helps to make the right decisions, e.g., about the appropriate method. **Machine learning, text mining, documentation, and reporting skills** are also mentioned as necessary skills for freelancers by 10 to 20%. Moreover, some interviewees (25%) note that freelancers should have an **algorithmic and logical way of thinking**, which includes breaking problems down into smaller parts and deduct, as well as maintain a big picture view throughout their work. Further, freelancers should be trustworthy, accurate, reliable, thorough, patient, and willing to learn new things. As such, those skills mirror the skills the interviewees ascribe to themselves.

### Difficulties with Outsourcing Data Analysis (RQ3)

Outsourcing data analysis tasks to freelancers is not necessarily an easy endeavor. Almost every interviewee is concerned about **communication issues** (80%). To begin with, a common language and shared understanding of the matter is necessary, which implies very clear requirements and well defined tasks. Since knowledge is sometimes assumed to be implicit and thus is not explicitly communicated between the freelancer and the customer, it can result in misunderstandings, inefficiencies, and in-transparency. Transparency (i.e., understanding of how the freelancer came to the results of data analysis) is crucial to guarantee the auditability of data (and the results). In addition, due to the distance, communication can take longer as one cannot simply go over to a colleague's desk but rather has to wait for a response (e.g., by email). Also, cultural differences may lead to communication problems, as noted by two interviewees. P2 noted: *"When you have foreign-language or diverse cultures, that maybe all speak English, but that maybe have a completely different understanding of a task, how to do it, and I would say the further away the cultures are from each other the more difficult it is."* The fact that information can be lost during intermediate steps of communication, called *"Chinese Whisper Effect"* by P8, is another possibly occurring problem. Additionally, the initial **briefing** is a related issue (30%): *"That's the tricky part, getting the briefing right"* (P4). Freelancers need to know exactly what they have to do, so the briefing has to be very precise, which means additional time and cost. This in turn raises the question if it is worth to outsource: *"Oftentimes you ask yourself, should I rather do it myself or train somebody locally, after all it's an investment"* (P14). **High setup costs and time**, not only regarding the briefing, but also the infrastructure, is mentioned by 15% as a barrier to outsourcing data analysis tasks to freelancers, and thus, the effort has to be *"justified"*, as stressed by P3. Connected to this issue is also the **knowledge gap** (40%) that results from the complex IT environment of a company and the entire domain knowledge that freelancers first have to familiarize with. Another big issue is **privacy** and confidentiality of data (55%). Similarly to other respondents, P8 says: *"The problem with outsourcing is always that you actually don't want to give away the data, primarily for us they are customer data."* This is also one of the reasons why some companies have not utilized outsourcing services so far and would feel uncomfortable sharing their data with freelancers. Ways to deal with this problem is anonymizing data or signing non-disclosure agreements. However, this problem still remains a sensitive issue. P6 points out the difficulty of finding a trade-off when he has to *"anonymize the sensitive data but to retain the utility of the data."* Furthermore, even if data is anonymized, it can happen that conclusions can be drawn about the actual identity of persons. Monitoring and **quality control** of work is another difficulty (40%). P5 says: *"If you were doing crowdsourcing, you have no guarantee, whatsoever, on the quality of the code or the piece of analytics that you get. So [...] somebody has to go and verify afterwards. And then it remains to be assessed [...], how much you benefit from crowdsourcing if you need to check afterwards what happened"*. **Trust** into freelancers (20%), the **vulnerability of data** whenever it is passed around (20%), the meeting of **deadlines** (15%), and the **danger of data manipulation** (15%), be it intentional or accidental, are other issues that have been mentioned several times. P20 asks himself: *"Does the guy have a huge incentive to disappear with the data and supply the competition with the analyses? Can I prosecute him? Can I find him at all?"* P13 sees the following danger with data analysis in general and even more with freelancers: *"I can always steer data analysis a little bit in a certain direction. So if you pursue some interests and know some statistics, I'm not saying cheating, but bending statistics in a way that they claim something even if it's not really true."* Thus, safeguards need to be applied. A problem that is not directly concerned with freelancers but rather with the company that is outsourcing data analysis is mentioned by 25% of interviewees: *"I think it's a huge problem that companies often don't even know what they can do in the first place"* (P3). Thus, they do not understand how they can draw meaningful insights through data science and hence have to be guided in the initial phase of exploring the possibilities of data analysis.

In addition, more than half of the interviewees (55%) have reported that their companies had outsourced data analysis to external specialized companies before. Also, they were asked if freelancers need to understand the structure and meaning of the data. Sixty-five percent answered to this question with "Yes", reasoning similarly to P9, who said: *"If you don't understand the meaning, then you will not be able to draw any meaningful conclusion."* Thirty-five percent answered that it would depend on the task and that if freelancers did not have to understand it, they would need very exact instructions, a *"cooking recipe"* (P18) to follow in order to execute the task. Furthermore, they were asked whether they would feel comfortable giving out their data to freelancers, for which 30% answered "Yes", 15% "No", and 55% with a mixed answer. Those whose answers were mixed argued that it depends on the data: With sensitive data or data concerning the core business they would not feel comfortable, whereas with anonymized or open access data they would agree doing so. Fifty-five percent of interviewees stated to have experienced problems in their job where freelancers could have supported in terms of time-related bottlenecks and lack of own resources or expertise.

### Analysis of Quantitative Data (RQ2)

We received 80 survey submissions from freelancers on Upwork and Freelancer, two most prominent OLM platforms. Respondents' experience with data analysis ranged from under a year to 45 years, with a median of 4 years. Noteworthy, many of the participants were beginners (i.e. freelancers with experience of just one year). Since data science is an emerging field, this could indicate that plenty of individuals are interested to start pursuing this profession. Another explanation can be the preference of beginners to seek for projects on OLM rather than through other recruitment channels. Participants rated their expertise on average with 3.82 and median of 4 on a five-point Likert scale. They stated to have learned about data analysis on the job (22%), through university courses (22%), through the Internet (17%), books (16%), online courses (15%), and teaching videos (8%). Most freelancers, 76%, were male, whereas 24% were female. 51% were between the age of 25 and 34, 24% between 18 and 24, and 13% between 35 and 44; the remaining 12% spread between 45 and over 65 years. Almost half of participants were living in European countries, a quarter in Asia and the remaining in America, Africa, or Australia. All participants either had a university degree or were enrolled as students. The level of education was, thus, relatively high, with 28% holding a doctorate, 40% a master's degree, and 25% a bachelor's degree. Their field of studies encompassed mainly Computer Science, Mathematics and Statistics, as well as Engineering. The majority, 59%, were employed in full or part-time jobs, 19% were currently looking for jobs, and 18% were students. Due to the high employment rate, 45% spent only less than 10 hours per week on freelance work. Furthermore, 19% spent up to 20 hours, 15% up to 30 hours, 6% up to 40 hours, and 15% more than that. Since our interviewees mentioned that freelancers should have expertise in the field the data analysis is conducted, we asked freelancers in what domains they were experienced ~~in~~. As a result, Mathematics & Statistics were chosen as the most common domains, followed by Science & Research, IT, Engineering, Business & Management, Economics, and Finance.

In order to test all skills from the survey for statistical significance we performed t-tests by comparing the mean values of the responses to the test value 3, which corresponds to having skills to a medium extent. Those items whose mean values were significantly larger or smaller than 3 were regarded as present, respectively absent on the platforms. Results show that all general skills, i.e. statistical, mathematical, database, programming, communication, presentation, visualization, machine learning, text mining, documentation and report writing skills, as well as data understanding were significantly different from 3. Thus, all mean values are larger, indicating that freelancers perceive their level of skills to be rather high, ranging from a mean of 3.28 in machine learning to a mean of 4.45 in data understanding (Table 1).

Performing the same tests for programming languages, tools, and data formats, we could obtain the following insights: Almost every item is significantly lower than 3. This implies that a lot of tools, programming languages, and data formats are not widely known by the surveyed freelancers. R, Python, and SQL, three items that were mentioned the most by our interviewees to be important for freelancers to have (90%, 65%, and 45% respectively), were found to be not significantly different from 3. This suggests on the one hand that these three tools are slightly better known than other tools, which had means lower than 3. But on the other hand, it also indicates that know-how for these tools is not highly present on freelancer platforms. Since particularly R and Python are some of the core skills required for data scientists (Kurgan and Musilek 2006), our findings support the fact that the widely discussed shortage of data scientists is also present on OLM (Harris et al. 2013). The only items having a statistically significant mean greater than 3 are Excel, PowerPoint, and CSV. Excel, with a mean of 4.21, is widely known among freelancers and they feel highly proficient in it. This is an important insight, as 40% of our interviewees stated Excel as a necessary skill to be known by freelancers. Again, we performed the same t-tests, this time for freelancers' knowledge of statistical methods. Even if freelancers are not very knowledgeable in statistical tools as indicated by the previous test, they state to have a high level of skills in statistical methods, particularly in descriptive and most inferential statistical methods. This is implied by the fact that most tested statistical methods have a mean value that is statistically significantly greater than 3. Machine learning techniques on the other hand have mean values lower than 3, partly statistically significant and partly not. This shows that these methods are not yet widely adopted by freelancers.

| Variable | N | T-test | | | Descriptive Statistics | | | | |
|---|---|---|---|---|---|---|---|---|---|
| | | t-value | p-value | mean ≠ 3 | min. | max. | mean | median | std. dev. |
| Data understanding | 80 | 21.116 | .000*** | Yes | 3 | 5 | 4.45 | 5 | .614 |
| Communication skills | 80 | 14.367 | .000*** | Yes | 2 | 5 | 4.23 | 4 | .763 |
| Documentation/Report skills | 80 | 11.075 | .000*** | Yes | 2 | 5 | 4.08 | 4 | .868 |
| Presentation skills | 80 | 9.786 | .000*** | Yes | 1 | 5 | 4.00 | 4 | .914 |
| Visualization skills | 80 | 9.649 | .000*** | Yes | 2 | 5 | 3.96 | 4 | .892 |
| Mathematical skills | 80 | 9.494 | .000*** | Yes | 1 | 5 | 3.91 | 4 | .860 |
| Statistical skills | 80 | 7.980 | .000*** | Yes | 1 | 5 | 3.83 | 4 | .925 |
| Programming skills | 80 | 6.749 | .000*** | Yes | 1 | 5 | 3.80 | 4 | 1.060 |
| Database skills | 80 | 3.717 | .000*** | Yes | 1 | 5 | 3.40 | 3 | .963 |
| Text Mining skills | 80 | 2.895 | .005* | Yes | 1 | 5 | 3.34 | 4 | 1.043 |
| Machine Learning skills | 80 | 2.085 | .040** | Yes | 1 | 5 | 3.28 | 3 | 1.180 |

Legend: ***: 0.001, **: 0.01, *: 0.05 significance level

**Table 1. T-test and Descriptive Statistics for General Skills**

To test whether correlation exists between any surveyed items, such as freelancers' experience, skills, and expertise in various tools, programming languages, and statistical methods, we performed Spearman rank sum correlations. As expected, the more years of experience freelancers have, the higher they rank their level of expertise in data analysis ($\rho$=0.603[**]). In turn, with increasing level of expertise they rate almost every general skill significantly higher (except database, programming, and machine learning skills), and almost all inferential statistical methods (excluding most of machine learning techniques). Interestingly, freelancers proficient in Python are also proficient in Machine Learning ($\rho$=0.51[**]), while those proficient in R are skilled in inferential statistics (($\rho$=0.537[**]). Even though these two topics overlap to some extent, this distinction can indicate the division between the tools traditionally used by ML experts and statisticians.

To compare interviewees' and freelancers' opinions on the obstacles that exist when outsourcing data analysis, we also asked freelancers open-ended questions about possible hurdles. One difficulty they see is that the problem has to be very well defined and requirements and specifications have to be clearly set (40%). Also, it has to be ensured that freelancers understand the problem and the goal of the analysis project (8.75%). Otherwise, as mentioned by one respondent, the following issue could arise: *"Freelancers might misunderstand the main objective of the project, thus building different models or using less-satisfactory techniques to solve the problem at hand."* In this regard, briefing (23.75%) is mentioned as an essential and also difficult phase of the outsourcing process, in which *"providing maximum information regarding the problem to be solved"* is necessary. Accordingly, communication is also mentioned by many participants (36.25%) as a difficulty when outsourcing data analysis to online freelancers. Knowledge gaps (13.75%) are also often identified obstacles in the outsourcing process, as e.g. mentioned by one respondent: *"Freelancers might not have the knowledge in the specific domain to conduct any meaningful interpretation of the results."* This is why it is even more important to find and choose freelancers that possess the necessary skills for a given project. They also need to be reliable (21.25%). Furthermore, quality of work is seen as a problem and, thus, appropriate monitoring and control needs to be applied (16.25%). Interestingly, confidentiality of data is stated as a problem only by 6 participants (7.5%), indicating that freelancers are not aware of this problem. Time zone differences, language barriers, and providing an accurate scope regarding time and price are seen as hurdles each by 5 participants.

## Discussion

Together, the qualitative and quantitative study results provide comprehensive information both about expected skills from freelancing data analysts and about the talent existing on major freelance platforms. Moreover, the interview results contribute to a better understanding of the obstacles towards outsourcing entire or parts of data analysis projects to OLM. Interestingly, the skills identified by the interviewed data scientists are not only limited to concrete skills picked up throughout studies such as math or coding (e.g. Kurgan and Musilek 2006), but go much beyond and include various skills required for data analysis. In the following, we discuss the answers to our previously stated research questions.

**(RQ1)** The most prominent skills data scientists should have, in accordance with literature, are *mathematical/statistical skills* and *technical affinity* such as database and programming capabilities. However, *in addition to those, our interviewees emphasized the importance of domain knowledge and communication skills for the success of an outsourced analysis project*. Also having an *eye for aesthetics* and details when visualizing data is a trait that is not necessarily associated with data science, but was mentioned by many of the interviewees as very important. Moreover, possessing the above mentioned skills does not immediately lead to being a good data scientist but, rather, *a combination of hard and soft skills*, understanding of data and knowing how to get the most out of it, altogether represent a good data scientist. This includes understanding the limits of what can be achieved with the data at hand and the ability to communicate those limitations to the clients. They in turn, according to the interviewed data scientists, do not have a thorough understanding of data analysis and see data science as an oracle, capable of answering any kind of questions. These excessively high expectations could be attributed to the spread of data science buzz to the mainstream of decision makers in recent years.

**(RQ2)** All skills that interviewees expected freelancers to have were statistically significant when tested, with means ranging from 3.28 to 4.45 on a 5-point Likert scale (Table 2). Therefore, concluding from the data, the *necessary skills to perform data analysis projects exist on freelance platforms and outsourcing them, or parts of them, to online freelancers is a feasible task with regard to the skills*. Table 2 is arranged in decreasing order of freelancers' self-reported skills (last column in the table). Interestingly, the most highly ranked skills are abilities attributed to the general data understanding, communication, and documentation - skills that are similar to so called "soft skills". This can be explained with the subjective character of these skills and might hint to the need to find additional approaches to evaluate these abilities. On the other hand, freelancers feel most unconfident about skills such as text mining and machine learning. This can be explained with long specialization required in order to be proficient in these topics. We also asked data scientists what skills they have in order to ascertain whether they project their own set of skills to those required from online freelancers or seek for experts with complementary skills (first column in the table). The *skills that data scientists expected from freelancers much more than they had themselves were documentation, visualization, and database skills*. Conversely, data scientists did not expect freelancers to be as good as they are in advanced knowledge that requires mathematics, statistics, and machine learning. It seems like *data scientists might be interested in workers with a complementary set of skills that could perform tasks which do not require advanced knowledge but rather skills that allow to perform general tasks such as extracting data or preprocessing*.

**(RQ3)** The success of outsourced data analysis projects is not only determined by the availability of required skills, but several other factors. Particularly, both interviewees and freelancers saw *communication issues as the biggest hurdle when outsourcing data analysis*. This includes the necessity to have clear requirements about the project, conducting precise briefings with the freelancer, establishing shared understanding of tasks, and maintaining good communication throughout the project. Also *quality assurance and knowledge gaps* are aspects that were mentioned by both interviewees and freelancers as hurdles in an outsourced project, whereas the latter laid even more emphasis on finding freelancers with appropriate knowledge and skills. *Privacy and confidentiality of data*, however, were mostly a concern of the interviewees; not as much by the surveyed freelancers. Hence, although outsourcing entails the hope to save resources in terms of time, money, and employees, outsourcing the project could require additional effort in terms of high setup costs and loss of time through additional communication, briefing, and performing quality assurance checks.

| Skills… | …Data scientists have | …Data scientists think that freelancers should have | …Freelancers have | |
|---|---|---|---|---|
| | | | Mean (1-5) | Std. Dev. |
| Data understanding | 45% | 50% | 4.45*** | 0.614 |
| Communication | 55% | 45% | 4.23*** | 0.763 |
| Documentation and writing | - | 10% | 4.08*** | 0.868 |
| Presentation | 15% | 10% | 4.00*** | 0.914 |
| Visualization | 10% | 40% | 3.96*** | 0.892 |
| Mathematics | 90% | 80% | 3.91*** | 0.86 |
| Statistical skills | 90% | 80% | 3.83*** | 0.925 |
| Programming | 75% | 65% | 3.80*** | 1.06 |
| Database | 55% | 80% | 3.40*** | 0.963 |
| Text Mining | - | 15% | 3.34** | 1.043 |
| Machine Learning | 30% | 20% | 3.28* | 1.18 |
| Domain Knowledge | 55% | 65% | Was asked directly | |
| Experience | 35% | 30% | Was inferred from years of experience and self-reported expertise on a 5-point scale | |

Legend: ***: 0.1%, **: 1%, *: 5% significance

**Table 2. Freelancer skills required according to data scientists compared with existing skills on freelance platforms**

Furthermore, interviewees and surveyed freelancers stated *preprocessing data as the most suitable task to outsource to online freelancers*. However, it is also the task that entails significant difficulties when outsourced. Additionally, data scientists noted data preprocessing to be the most tedious task and the one that they would like to outsource. Despite the various obstacles in outsourcing this step they would be eager to see solutions that would allow to overcome these obstacles and therefore reduce their workload. *Problems identified during the interviews for the data cleaning process* were: (1) possibly confidential data, (2) the necessity of domain knowledge to understand the data, (3) the fact that data cleaning entails many tacit, subjective assumptions, and (4) the necessity of putting significant trust into freelancers when handing them data. Moreover, (5) data cleaning is a complex process, which requires a lot of customer contact and has to be repeated iteratively. Hence, the (6) coordinating effort with the freelancer is significant, as it has to be constantly maintained, and (7) specifications and results have to be clearly conveyed. The next most mentioned task suitable for outsourcing to online freelancers, as mentioned by the interviewees, is *data collection*. Difficulties that are likely to arise are due to data spread over various sources such that freelancers first need to gain an understanding where the needed data is stored and how to get access to it. Also, data collection is error-prone and needs to be auditable. Additionally, several interviewees preferred to *outsource only the entire data analysis process*, since all steps within the project are connected: breaking them apart would lead to loss of knowledge and complicate the project. Problem definition and project specification were argued to be difficult to outsource by both interviewees and freelancers, as they require a lot of knowledge about the company, the domain, and the data. Interestingly, opinions about outsourcing data modeling were divided equally among interviewees. Again, knowledge about domain, background, and data were mentioned as obstacles in addition to the need to be able to handle large data, which require a lot of storage and computer power.

The needs-supplies conceptualization of the Job-Fit theory provides us with theoretical justification to examine the skills required for online data analysis (i.e., the needs) compared to the existing pool of talents (i.e., the supplies) on OLM. However, to support future theorization on a job's fit to OLMs it is necessary to better understand the skills and abilities required in this setting. Our study takes the first step towards reaching this goal through thorough methodological exploration of the skills required for freelance data analysis. We chose data analysis as a prototypical job of online labor markets for two reasons: (1) it is a common domain in all major OLMs and (2) data science is a sought after domain with a substantial shortage of experts. Consequently, *by focusing on the data analysis domain our study additionally has potential to inform practitioners by shedding light on the existing talent on OLM and discussing hurdles for outsourcing data analysis to such platforms*.

## Limitations and Future Research

It is important to note that our approach has the following limitations. First, we have conducted an exploratory study, mainly bounded to the explicit factors related to the needs-supplies conceptualization of Job-Person fit. However, behavioral factors such as cognitive abilities, personal desires or satisfaction, and other psychological needs were not considered. Hence, future work will have to examine topics related to the behavioral specifications of candidates. Second, this study's cross-sectional design could be expanded to a longitudinal design to explore how abilities, skills, and expertise develop over time. Third, our study relies on the possible biased self-reported skills by the freelancers. Alternatively, one might want to assess the freelancers' abilities through tests. Whilst tests would be a more reliable approach to assess a *given* list of skills our design aimed at collecting comprehensive data from a big sample of freelancers active on different major OLM platforms to elicit an initial overview of the skills. Future research might sacrifice breadth and generalizability over different platforms for the sake of an unbiased assessment. Lastly, the geographical proximity of interviewees and their residence in LOCATION SUPRESSED could be a potential argument against generalizability. We tried to address this limitation by interviewing a relatively large number of experts and ensuring that most of them work in international companies. Future studies will, of course, have to confirm the absence of a locality bias.

## Conclusion

Most organizations experience an alarming shortage of data analysis experts in an emerging world of omnipresent data. To our knowledge *this study represents the first methodological attempt to explore the potential of overcoming the shortage of data analysis experts by outsourcing the analysis, or parts of it, to freelancers available on OLMs*. Specifically, we first explored the skills required for data analysis in general and then elicited the skills expected from freelancers on OLM from data scientists. We further investigated the obstacles to data analysis outsourcing and possible remedies to overcome them. As it has been outlined in the introduction, this study adopted the needs-supplies conceptualization of the Person–Job fit theory and focused on the required skills, knowledge, and expertise. The results presented here can be useful for future theorization on employee selection in online settings and serve as a starting point for extending the scope from data analysis to other areas common on online labor markets. The results of this study demonstrate that *the skills required for data analysis exist on major freelance platforms and that outsourcing data science projects, or parts of them, to OLMs is feasible*. These skills include data understanding, communication, documentation, presentation, visualization, and mathematical/statistical and technical abilities like programming, database, text mining, and machine learning. Furthermore, although data analysis outsourcing faces various hurdles (e.g., communication issues, knowledge gaps, quality of work, or data confidentiality), this study provides some evidence that they can be resolved, thus, making outsourcing data analysis tasks possible. As such, it highlights a possible approach for overcoming the scarcity of data science professionals.